\documentclass[12pt]{article}
\usepackage{graphicx}
\begin{document}
\rightline{NKU-06-SF1}
\bigskip

\begin{center}
{\Large\bf Thermodynamics of Born-Infeld-anti-de Sitter  black holes in the grand canonical ensemble }

\end{center}
\hspace{0.4cm}
\begin{center}
{Sharmanthie Fernando \footnote{fernando@nku.edu}}\\
{\small\it Department of Physics \& Geology}\\
{\small\it Northern Kentucky University}\\
{\small\it Highland Heights}\\
{\small\it Kentucky 41099}\\
{\small\it U.S.A.}\\

\end{center}

\begin{center}
{\bf Abstract}
\end{center}

\hspace{0.7cm}{\small

The main objective of this paper is to study  thermodynamics and stability of static electrically charged Born-Infeld black holes in AdS space in $D=4$. The Euclidean action for the grand canonical ensemble is computed with the appropriate boundary terms. The  thermodynamical quantities such as the Gibbs free energy, entropy and specific heat of the black holes are derived from it. The global stability of black holes are studied in detail by studying the free energy for various potentials. For  small values of the potential, we find that there is a Hawking-Page phase transition between a BIAdS black hole and the thermal-AdS space. For large potentials, the black hole phase is dominant and is preferred over the thermal-AdS space. Local stability is studied by computing the specific heat for constant potentials. The  non-extreme black holes have two branches: small black holes are unstable and the large black holes are stable. The extreme black holes are shown to be stable both globally as well as locally. In addition to the thermodynamics, we also show that the phase structure relating  the mass $M$ and the charge $Q$ of the black holes is similar to the liquid-gas-solid phase diagram. 
}

\section{Introduction}

Hawking and Page showed   the existence of a phase transition between AdS black holes and thermal AdS space \cite{page}. In extending these concepts, it was shown that such phase transitions are not unique to the AdS spaces, but may occur in asymptotically dS spaces and flat spaces as well \cite {carlip}. On the other hand, in recent years, thermodynamics of black holes in AdS space has generated renewed attention due to the AdS/CFT duality \cite {mal}. Such a duality relate thermodynamics of black holes in AdS space to the thermodynamics of dual CFT. Hence by studying the phase transitions of AdS black holes leads to  a novel way  of studying  phase transitions in the dual field theories \cite{witten}.

In this paper our main  focus is on thermodynamics of black holes  in Einstein-Born-Infeld gravity in AdS space.  Born-Infeld electrodynamics was first introduced in 1930's to obtain a classical theory of charged particles with finite self-energy \cite{born}. Born-Infeld theory has received renewed attention since it turns out to play an important role in string theory. It arises naturally in open superstrings and in D-branes \cite{leigh}. The low energy effective action for an open superstring in loop calculations lead to Born-Infeld type actions \cite{frad}. It has also been observed that the Born-Infeld action arises as an effective action governing the dynamics of vector-fields on D-branes \cite{tsey1}. For a  review of aspects of Born-Infeld theory in string theory see Gibbons \cite{gib1} and Tseytlin \cite{tsey2}.

One of the interesting features of Born-Infeld electrodynamics is that the electric field for a point particle is $E = Q/\sqrt{ r^4 + Q^2/b^2}$ leading to a screening effect on $E$ for small values of $r$. Here, $b$ is the  non-linear parameter in the theory. Hence one can interpret the point-particle object as an extended source with an effective radius $r_o = Q^2/b^2$. Lattice simulations of Bonn-Infeld-QED was done in \cite{sinclair} and the short distance screening of $E$ was shown to be similar as in  the classical theory. When Born-Infed electrodynamics is coupled to gravity, the above behavior makes drastic changes to the singular nature of the black hole solutions of the theory as compared to the ones resulting when gravity is coupled to Maxwell's electrodynamics. Such properties are discussed in detail in section 2 of this paper. 

The Born-Infeld black hole with a zero cosmological constant was obtained by Garcia et.al.\cite{garcia} in 1984. Two years later, Demianski \cite{demia} also presented a solution with the title ``Static Electromagnetic Geon'' which  differs with the one in \cite{garcia} by a constant. Trajectories of test particles in the static charged Born-Infeld black hole was discussed by Breton \cite{nora1}. This black hole in isolated horizon framework was discussed by the same author in \cite{nora2}. Gibbons and Herdeiro \cite{gib2} derived a Melvin Universe type solution describing a  magnetic field permeating the whole Universe in Born-Infeld electrodynamics coupled to gravity. By the use of electric-magnetic duality, they also obtained Melvin electric and dyonic Universes. Static particle-like and black hole solutions for the Einstein-Born-Infeld-dilaton system were constructed in \cite{tamaki1} \cite{clem}. Black hole solutions to the theory with derivative corrections to the Born-Infeld action were derived in \cite{tamaki2}.

In this paper, we  focus on the static charged black holes of the Born-Infeld electrodynamics in AdS space. This is the non-linear generalization of the Reissner-Nordstrom black hole (RNAdS) and is characterized by charge $Q$, mass $M$ and  the non-linear parameter $b$. Such black holes have been presented in \cite{fer1} \cite{cai} \cite {dey}. As a new result, we present an interesting phase structure between the mass $M$ and the charge $Q$ of the BIAdS black holes. We will also derive the thermal properties of the black holes from the path-integral approach developed by Gibbons and Hawking \cite{gibbons4}. In this approach, the Euclidean partition function for quantum gravity is interpreted as the thermal partition function for a given ensemble with the black hole temperature $T$. We will compute the Euclidean action for black holes in the grand canonical ensemble where the electrostatic potential at infinity is fixed. The pure-AdS space is treated as the background space for this computation. Corresponding thermodynamic quantities such as Gibbs free energy, entropy and specific heat are calculated. The global and local thermodynamic stability of the BIAdS black holes are studied in detail. Some thermal properties of BIAdS and BIdS black holes  were studied in \cite{dey} \cite{cai}. First law of thermodynamics for BI black holes was discussed in \cite{ras}. However, there are new results related to these black holes, which were not presented in the above papers.

The paper is presented as follows: In section 2, the Born-Infeld black hole solutions are discussed. In section 3, the Euclidean action is computed. Thermodynamics and stability are discussed in  section 4. Finally, the conclusion is given in section 5.

\section{Born-Infeld black holes in anti-de Sitter space}

In this section we will present details of the static spherically symmetric charged black holes in Born-Infeld non-linear electrodynamics coupled to gravity in $D=4$. These black holes have been discussed in \cite{fer1} \cite{cai} \cite{dey}. However, there are interesting details to the properties of the singularities and the horizons when the parameters of the black hole is varied which were not  discussed in the above papers which will be highlighted here.

The Einstein-Born-Infeld action in $D=4$ is given by,
\begin{equation}
S =  \frac{1}{ 16 \pi G} \int d^4x \sqrt{-g} \left[ (R - 2 \Lambda)  + L(F) \right]
\end{equation}
where,
\begin{equation}
L(F) = 4 b^2 \left( 1 - \sqrt{ 1 + \frac{ F^{\mu \nu}F_{\mu \nu}}{ 2 b^2}} \right)
\end{equation}
Here, $b$ is the Born-Infeld parameter and $\Lambda = - 3/l^2$ is the negative cosmological constant. The parameter $b$ is related to the sting tension $\alpha'$ as $ b = 1/(2 \pi \alpha')$. Note that in the limit $ b \rightarrow \infty $, $L(F)$ takes the form, 
\begin{equation}
L(F) = - F^{\mu \nu}F_{\mu \nu} + O(F^4)
\end{equation}
reducing to the standard Maxwell's form.

By solving the equations of motion, the Born-Infeld-anti-de Sitter (BIAdS) black hole solutions can be written as,

\begin{equation}
ds^2 = - f(r) dt^2 + f(r)^{-1} dr^2 + r^2  d \Omega^2
\end{equation}
with,
\begin{equation}
f(r) = 1 - \frac{2M}{r} + \frac{r^2}{l^2} + \frac{2 b r^2}{3} \left( 1 - \sqrt{ 1 + \frac{Q^2}{r^4 b^2}} \right) +  \frac{ 4 Q^2}{ 3 r} H(r) 
\end{equation}
The function $H(r)$ is given by,
\begin{equation}
H'(r) =  \frac{ -1}{ \sqrt{ r^4 + \frac{Q^2}{b^2}} }
\end{equation}
To obtain the black hole solution, the above equation can be integrated as,
\begin{equation}
H(r) = \int^{\infty}_{r}  \frac{ d \tilde{r} }{ \sqrt{ \tilde{r}^4 + \frac{Q^2}{b^2} } } =  \sqrt{ \frac{b}{ 4 Q} } { \cal K }\left( Cos^{-1} ( \frac{ r^2 - Q/b} { r^2 + Q/b} ), \frac{1}{\sqrt2} \right)
\end{equation}
Here ${\cal K }$ is an elliptic integral of the first kind \cite{abra}. The elliptic integral also can be written as a Hypergeometric function ${\cal F}$ with the relation,
\begin{equation}
\frac{ r}{ 2 Q} {\cal K} \left( Cos^{-1} \left( \frac{ r^2 - \frac{Q}{b}} { r^2 + \frac{Q}{b}} \right), \frac{1}{\sqrt2} \right)
  =  {\cal F } \left( \frac{1}{4}, \frac{1}{2}, \frac{5}{4}, -\frac{Q^2}{ b^2 r^4} \right)
\end{equation}
Replacing the $H (r) $ in the eq.(5) by the Hypergeometric function one obtains,
\begin{equation}
f(r) = 1 - \frac{2M}{r} + \frac{r^2}{l^2} + \frac{2 b r^2}{3} \left( 1 - \sqrt{ 1 + \frac{Q^2}{r^4 b^2}} \right) + \frac{ 4 Q^2}{ 3 r^2} \hspace{0.2cm}  {\cal F} \left( \frac{1}{4}, \frac{1}{2}, \frac{5}{4}, -\frac{Q^2}{ b^2 r^4} \right)
\end{equation}
This is the form obtained in the references \cite{cai} \cite{dey}. We will continue to use this form in the rest of the paper. The black hole solutions without the cosmological constant was obtained by Garcia et.al. in \cite{garcia}. Note that there is another solution with,
\begin{equation}
H(r) = - \int^{r}_{0}  \frac{ d \tilde{r} }{ \sqrt{ \tilde{r}^4 + \frac{Q^2}{b^2} } } = -  \sqrt{ \frac{b}{ 4 Q} } { \cal K }\left( Cos^{-1} \left( \frac{\frac{Q}{b} -  r^2} { \frac{Q}{b} + r^2 } \right), \frac{1}{\sqrt2} \right)
\end{equation}
Here, the integration limits differ to the one in eq.(7). The $H(r)$ obtained in this case has the property that $f(r)$ is regular at $r=0$ when $ M =0$. For $ \Lambda =0$, such  solutions were presented as ``Static electromagnetic geons'' by Demianski \cite{demia}. The difference between these two solutions is explained in detail in \cite{nora3}.

The electric field $F_{tr} =E$ for the black hole solution is given by,
\begin{equation}
E(r)  = \frac{Q}{\sqrt{ r^4 + \frac{Q^2} {b^2}}}
\end{equation}
The electric gauge potential is given by,
\begin{equation}
A_t = \frac{Q}{r}  {\cal F} \left( \frac{1}{4}, \frac{1}{2}, \frac{5}{4}, -\frac{Q^2}{ b^2 r^4} \right) -  \Phi
\end{equation}
where $\Phi$ is a constant. In order to fix the gauge potential at the horizon to be zero, i.e. $A_t(r_+) =0$, the quantity $\Phi$ is chosen as,
\begin{equation}
\Phi = \frac{Q}{r_+}  {\cal F} \left( \frac{1}{4}, \frac{1}{2}, \frac{5}{4}, -\frac{Q^2}{ b^2 r_+^4} \right)
\end{equation}
Hence, $\Phi$ represents the electrostatic potential difference between the horizon and the infinity. $\Phi$ will play an important role in section 3 when the thermodynamical stability is analyzed.

The behavior of the function $f(r)$ closer to the origin is  important in observing variety of solutions with differing singular structure for the BIAdS black holes. Hence for small $r$, 
\begin{equation}
f(r) \approx 1 - \frac{( 2M - a)}{r} - 2 b Q + \frac{2 b^2}{3} r^2  + \frac{r^2}{l^2} + \frac{ b^2}{5}  r^4
\end{equation}
Here,
\begin{equation}
a = \sqrt{ \frac{b}{ \pi} } Q^{3/2} \Gamma \left(\frac{1}{4} \right)^2
\end{equation}
The nature of the curvature singularity at $r=0$ depends on the relation between $ M $ and  $a/2$. When $ M > a/2$, the space-time is similar to the Schwarzchild-anti-de-Sitter (SchAdS) black holes. When $ M < a/2$, the space-time is similar to the RNAdS black holes. In this case it is possible to have zero, two or one degenerate horizon. When $M = a/2$, the function $f(r) = 1 - 2 b Q$ at $r =0$ which is finite. However, some curvature invariants do diverge at $r=0$ for this case \cite{nora2}.

In the limit $b \rightarrow \infty$, the elliptic integral in eq.(7) can be expanded to give,
\begin{equation}
f(r)_{RNAdS} = 1 - \frac{2 M}{r} + \frac{ Q^2}{r^2} + \frac{r^2} {l^2}
\end{equation}
resulting  in the function $f(r)$ for the Reissner-Nordstrom-AdS (RNAdS) black hole for Maxwell's electrodynamics.

\subsection{Extreme Black Holes}

To further our understanding as to how the parameters of the black holes characterize the nature of the singularities, the extreme black hole solutions will be discussed here. Note that for extreme black holes, both $f(r)$ and $df(r)dr$ has to be zero at the horizon. Such conditions will lead to the equation,
\begin{equation}
1 + \left( 2 b^2 + \frac{3}{l^2} \right) r_{ex}^2 - 2 b \sqrt{ r_{ex}^4 b^2 + Q^2} =0
\end{equation}
which gives the solution as,
\begin{equation}
r^2_{ex} = \frac{ ( 2 b^2 + \frac{3}{l^2}) + \sqrt{\delta }}{ \frac{3}{l^2} ( 4 b^2 + \frac{3}{l^2})}
\end{equation}
Here,
\begin{equation}
\delta = \left(2 b^2 + \frac{3}{l^2}\right)^2 + ( 1 - 4 b^2 Q^2) \left( 4 b^2 + \frac{3}{l^2}\right) \frac{3}{l^2} 
\end{equation}
There will be a real root for $r_{ex}$ only when $ 1 - 4 b^2 Q^2 < 0$. Hence even if the solutions are of RNAdS type, there will be two horizons only when $Q b > 1/2$. 
The mass of the extreme black hole  is given by,
\begin{equation}
M_{ex}  =   \frac{r_{ex}^2}{3} + \frac{ 2 Q^2}{ 3 r_{ex}} \hspace{0.2cm}   {\cal F} \left( \frac{1}{4}, \frac{1}{2}, \frac{5}{4}, -\frac{Q^2}{ b^2 r_{ex}^4} \right)
\end{equation}
If $ M > M_{ex}$ then there will be two horizons. For $ M = M_{ex}$ there will be a degenerate horizon. If $M < M_{ex}$, there are no  horizons and will yield a naked singularity. Note that these three cases can be discussed only for  $ M < a/2$ leading to RNAdS type solutions. If $ M > a/2$ there will be a horizon always  without the possibility of extreme black holes. Hence one can give a general description of the nature of the solutions for all the  bounds discussed above as follows;
\newline
\newline
If $ M > a/2 $  $\Rightarrow$ SchAdS type black hole with one horizon
\newline
If $M = a/2$ and $ Q b  >1/2$ $\Rightarrow$ the solution is SchAds type black hole with one horizon
\newline
If $ M = a/2$ and $ Q  b < 1/2$ $\Rightarrow$ the solution is RNAdS type naked singularity
\newline
If  $ M_{ex} < M < a/2 $    $\Rightarrow$ RNAdS type black hole with two horizons.
\newline
If $ M < M_{ex} < a/2 $   $\Rightarrow$ RNAdS type Naked singularity.
\newline
If $ M = M_{ex} < a/2 $ $\Rightarrow$ RNAdS type extreme black hole with a degenerate  horizon
\newline
\newline
These categories of solutions can be well represented by the following graphs.

\vspace{0.3cm}
\begin{center}
\scalebox{.9}{\includegraphics{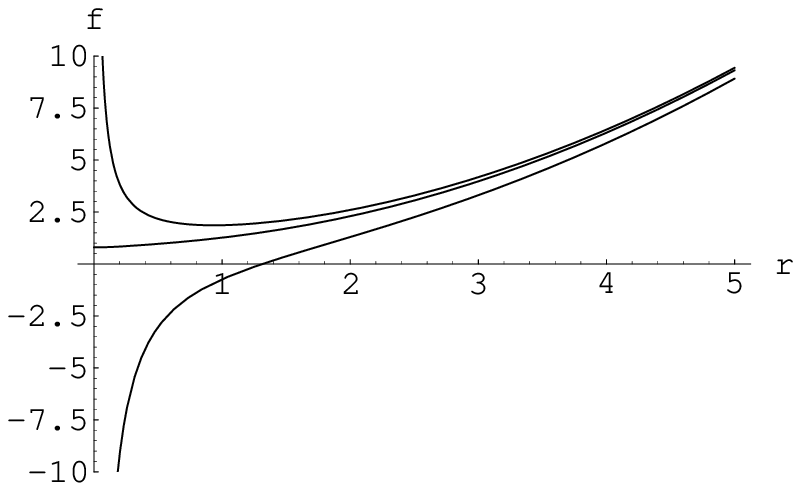}}
\vspace{0.3cm}
\end{center}
FIG.1: The function $f(r)$ vs $r$ for    $ Q b < 1/2 $. Here $b = 1$, $ Q=0.1  $,  $ l = \sqrt{3}$ and $ G =1$ 
\vspace{0.3cm}
\begin{center}
\scalebox{.9}{\includegraphics{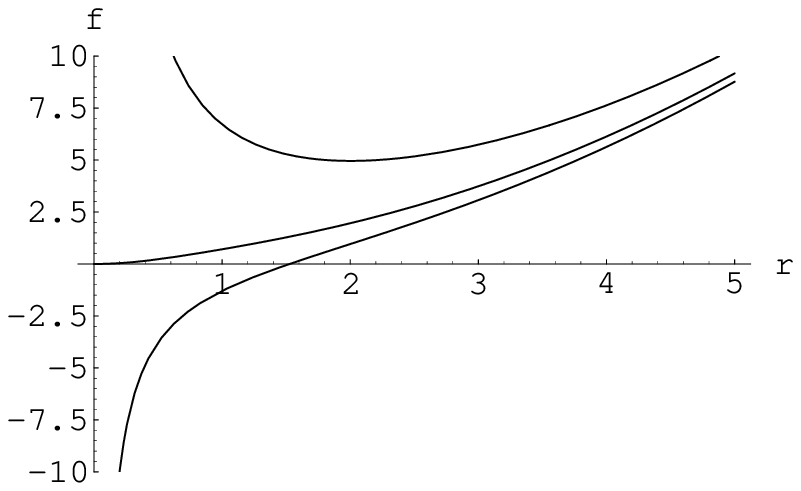}}
\vspace{0.3cm}
\end{center}
FIG.2: The function $f(r)$ vs $r$ for    $ Q b = 1/2 $. Here $b = 1$, $ Q= 0.5 $,  $ l = \sqrt{3} $ and $ G =1$ 
\vspace{0.3cm}
\begin{center}
\scalebox{.9}{\includegraphics{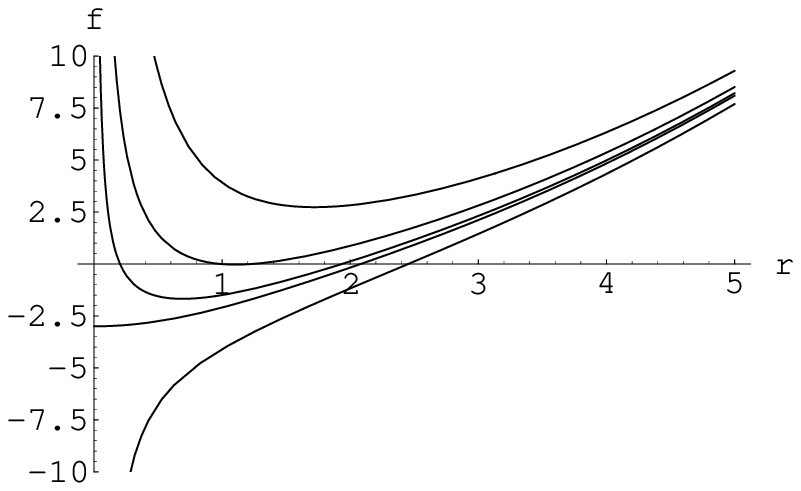}}
\vspace{0.3cm}
\end{center}
FIG.3: The function $f(r)$ vs $r$ for    $ Q b > 1/2 $. Here $b = 1$, $ Q= 2  $,  $ l = \sqrt{3}$ and $ G =1$ 
\newpage
All these categories can be represented in one graph for $M$ vs $Q$ for fixed $b$ value as in the following figure.
\vspace{0.3cm}
\begin{center}
\scalebox{.9}{\includegraphics{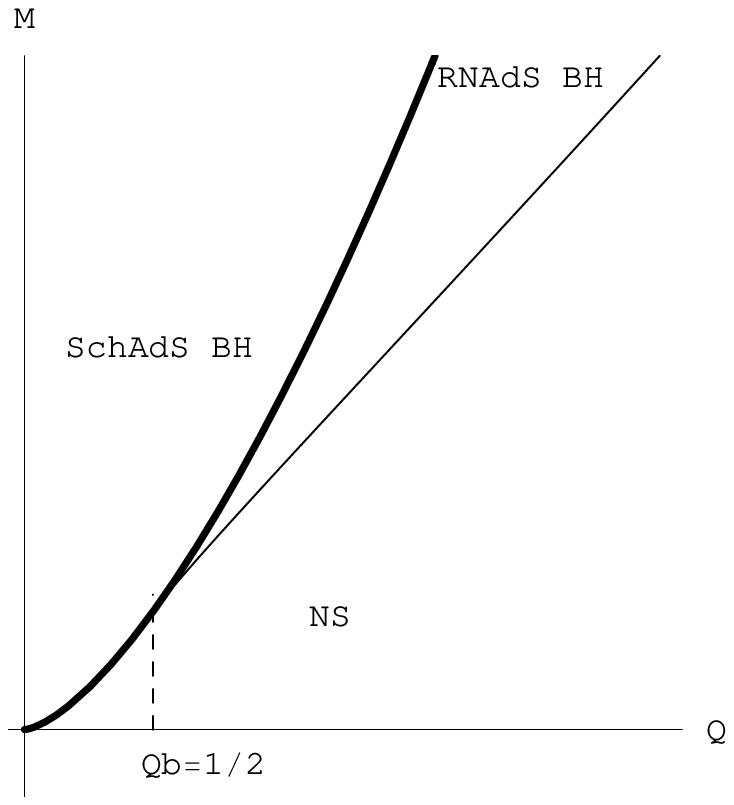}}
\vspace{0.3cm}
\end{center}
FIG.4: The function $M$ vs $Q$ for    $ b = 0.25 $. Here $ l = \sqrt{3000}$ and $ G =1$. The dark line shows the $M = a/2$ bound and the light line shows the $M = M_{ex}$ bound. ``NS'' represents Naked Singularities. 
\newline
\newline
The above diagram reminds us the  liquid-gas-solid phase diagram with $M$  representing the pressure $P$ and $Q$ representing the temperature $T$ \cite{phase}. The SchAdS type black holes are like the solid phase, the RNAdS type black holes are like the liquid phase and the Naked Singularities are like the gaseous phase. The point  where all three types of solutions meet in the figure represents the ``Triple Point'' where all three phases co-exists. This point is located at $Q = \frac{1}{2b}$ and $M= 
\frac{a}{2} = \frac{1}{2} \sqrt{ \frac{b}{  \pi} } Q^{3/2} \Gamma \left(\frac{1}{4} \right)^2
=
\frac{1.31103}{b}$. The non-linear nature of the Born-Infeld theory has led to the possibility of  a variety of solutions, which would not be possible in the RNAdS black hole case. Having a variety of solutions has led to this interesting phase structure.

In comparison, the RNAdS black holes corresponding to Maxwell's electrodynamics do have an extreme black holes with the radius $r_{ex}$ at
\begin{equation}
r_{ex}^2 = \frac{1}{6} ( -l^2 + l \sqrt{ l^2 + 12 Q^2} )
\end{equation}
The mass of the extreme black hole is,
\begin{equation}
M_{ex} = \frac{r_{ex}}{2} + \frac{ r_{ex}^3 } { 2 l^2} + \frac{ Q^2}{ 2 r_{ex} }
\end{equation}
Depending on whether $ M > M_{ex}$, the solution will have two horizons, one degenerate horizon or none. The following figure represents the possible cases. It must be obvious how the non-linear nature brings out a rich phase structure to the BIAdS solutions as compared to the RNAdS solutions.

\vspace{0.3cm}
\begin{center}
\scalebox{.9}{\includegraphics{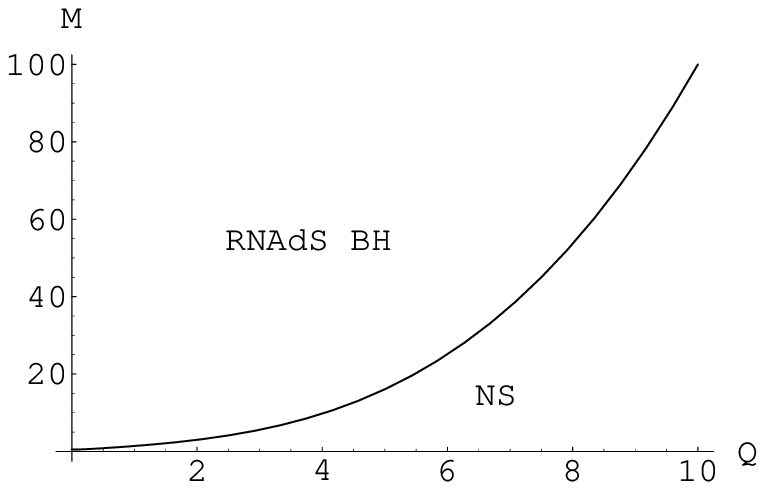}}
\vspace{0.3cm}
\end{center}
FIG.5: The function $M$ vs $Q$ for   RNAdS black holes. Here $ l = 1$ and $ G =1$. The graph is the $M=M_{ex}$ bound.

\section{Euclidean action calculation}

In this section we will compute the Euclidean-Born-Infeld action which facilitate the study of thermodynamics of black holes in the grand canonical ensemble. The procedure for computing the Gibbs-Euclidean action is similar to the one initiated by Hawking and Page \cite{page} for the SchAdS black hole in $D=4$. Extension of this work applied to the RNAdS black holes in all dimensions is given in \cite{empa}.

In this approach, initially the electrical potential and the temperature is fixed on a boundary with a finite radius $r_B$. The Euclidean action has three components; bulk, surface and the counter term. The counter term comes from the pure $AdS_4$. It is necessary to  have a counter term since the integrals of the bulk and the surface terms  diverge when $r_B$ is sent to infinity at the final stage of the computation to remove the artificial boundary. Once the Euclidean action is computed one can obtain the Gibbs free energy, which will facilitate the study of  the global stability of the black holes.

First, to make the action Euclidean, the time coordinate $t$ is substituted with $i \tau$. This  will  make the metric positive definite:
\begin{equation}
ds^2 =  f(r) d \tau^2 +  f(r) ^ {-2} dr^2 + r^2 d \Omega^2
\end{equation}
There is a conical singularity at the horizon $ r = r_{+}$ in the Euclidean metric. To eliminate it, the Euclidean time $\tau$  is made periodic with period $\beta$. Here, $\beta = 1/T$ where $T$ is the Hawking temperature. The Hawking temperature of the black hole solutions discussed above can be calculated as follows: $T = {\kappa}/2 \pi$. Here $\kappa$ is the surface gravity given by 
$$ \kappa = - \frac{1}{2} \frac{f(r)}{dr} |_{r= r_{+}} $$
Here, $r_+$ is the event horizon of the black hole. Since $f(r) = 0$ at $r = r_+$, the eq.(5) can be used to calculate the surface gravity exactly and the corresponding temperature as,
\begin{equation}
T = \frac{1}{4\pi} \left( \frac{1}{r_+} - \Lambda r_+ + 2 b^2  r_+  - \frac{2 b \sqrt{Q^2 + r_+^2 b^2}}{r_+}  \right)
\end{equation}
Now the Euclidean Einstein-Born-Infeld action can be written as,
\begin{equation}
I =  - \frac{1}{ 16 \pi G} \int_{{\cal M}}  dx^4 \left( R - 2 \Lambda + L(F) \right) - \frac{1}{ 8 \pi} \int_{ \partial {\cal M}} dx^3 \sqrt{h} K  + I_{counter term}
\end{equation}
The above action is written on a compact region given by ${\cal M}$ and the boundary $\partial {\cal M} $. The boundary is at a finite radius $r = r_B$ and it has the topology $ S^1 \times S^2 $. Here $h_{ij}$ is the induced metric on the boundary and, $K$ is the trace of the intrinsic curvature of the boundary.

The first term is the action defined on the bulk. The second term defined on the boundary is the Gibbons-Hawkings boundary term \cite{gibbons4}. Note that since gauge potential is fixed at infinity, the boundary terms related to the gauge field will not contribute to the Euclidean action. To obtain a finite value for the Euclidean action one has to introduce a  counter term. In this case the counter term will be the minus of the Euclidean action for the pure-AdS geometry. Note that there are other approaches to compute the counter terms  \cite{bala} \cite{mann}  \cite{empa2}  \cite{rod}. Choosing an appropriate reference background to compute the counter terms has its own complications as discussed in \cite{empa}.

The equations of motion derived from the action of the theory in eq.(1) is,
\begin{equation}
R_{ab} - \frac{1}{2} g_{ab} R + \Lambda g_{ab} =  T_{ab}
\end{equation}
\begin{equation}
\bigtriangledown _a \left(  \frac{F^{ab}}{ \sqrt{ 1  + F^2} } \right) = 0
\end{equation}
The scalar curvature $R$ can be obtained from eq.(26) as,
\begin{equation}
R = 4 \Lambda - T
\end{equation}
where,
\begin{equation}
T =  \frac{2 F^2} { \sqrt{ 1 + F^2} } + 2 L(F)
\end{equation}
Substituting the above relations, one can compute  the on-shell bulk contribution $I_{bulk}$ as,
\begin{equation}
I_{bulk} = - \frac{1}{ 16 \pi G} \int _{M} dx^4  \sqrt{g}  \left( 2 \Lambda -  L(F) - \frac{2 F^2} { \sqrt{ 1 + F^2}} \right)
\end{equation}
Note that in the limit $ b \rightarrow \infty $, the above action becomes,
\begin{equation}
I_{bulk RNAdS} = - \frac{1}{ 16 \pi G} \int _{\cal M} dx^4  \sqrt{g}  ( 2 \Lambda -  F^2 )
\end{equation}
which is  the same as given in \cite{empa}. Substituting for the electric field  $F_{r \tau }$ given by,
\begin{equation}
F_{ r \tau} = i F_{rt} = \frac{ i Q}{ \sqrt{ r^4  +  \frac{ Q^2}{b^2} }}
\end{equation}
$I_{bulk}$ can be integrated to be,
\begin{equation}
I_{bulk} =  \frac{ \omega \beta} { 16 \pi G} \int^{r_B} _{r_+}  \left( \frac{ 6 r^2}{ l^2} +  4 b^2  r^2   - 4 b \sqrt{ r^4 b^2 + Q^2} \right) dr
\end{equation}
Here $\omega$ is the volume of the two-sphere and $\beta$ is the inverse temperature. The integration with respect to $r$ yields,
\begin{equation}
I_{bulk} = \frac{ \omega \beta}{ 16 \pi G}  \left[ \frac{ 2}{l^2}  ( r_B^3 - r_+^3 )  + \frac{ 4 b^2} { 3} ( r_B^3 - r_+^3)  - 4 b \int_{r_{+}}^{r_B} \sqrt{ r^4  + Q^2 } \right]
\end{equation}
The Euclidean action for the pure AdS space is given by,
\begin{equation}
I_{AdS} =  \frac{ \omega \beta_0} { 16 \pi G} \int^{r_B}_{r_{+}}   \frac{ 6 r^2 } { l^2} 
\end{equation}
Since we are computing this for $D=4$ which is an even dimension, $ r_+ =0$ for the pure-AdS space \cite{gibbons2}. For odd dimensions, $r_+$ is not zero for pure-AdS space-times as discussed in \cite{gibbons2}. Note that $\beta_0$ is the  time period for pure AdS case. It has to be rescaled to match with the period $\tau$:
\begin{equation}
f(r) \tau^2 = \tau_0^2 \left(  1 + \frac{r^2}{l^2} \right)
\end{equation}
After some approximations,
\begin{equation}
\beta_0 = \beta \left(  1 - \frac{ M l^2} {r^3} 
+ \frac{2 b^2 l^2}{3} \left( 1 - \sqrt{ 1 + \frac{Q^2}{r^4 b^2}} \right) + \frac{ 2 Q^2 l^2}{ 3 r^4}   {\cal F} \left( \frac{1}{4}, \frac{1}{2}, \frac{5}{4}, -\frac{Q^2}{ b^2 r^4} \right) \right)
\end{equation}
The final result for the Euclidean-Born-Infeld action can be obtained by sending $r_B$ to infinity leading to the   following value,
\begin{equation}
I_E = \frac{ \omega \beta} { 16 \pi G } \left[ r_+  - \frac{r_+^3}{l^2} -   \frac{2 r_+^3 b^2}{3}  \left( 1  -   \sqrt{ 1 + \frac{Q^2}{ r_{+}^4 b^2} } \right) -\frac{ 4 Q^2}{ 3 r}  \int^{\infty}_{r}  \frac{ d \tilde{r} }{ \sqrt{ \tilde{r}^4 + \frac{Q^2}{b^2} } } \right]
\end{equation}
In deriving this we have made use of the fact that $f(r_+) = 0$. One can replace the integral with the Hypergeometric function $\cal F$ which brings the action to be,
\begin{equation}
I_E = \frac{ \omega \beta} { 16 \pi G } \left[ r_+  - \frac{r_+^3}{l^2} -   \frac{2 r_+^3 b^2}{3}  \left( 1  -   \sqrt{ 1 + \frac{Q^2}{ r_{+}^4 b^2} } \right) -\frac{ 4 Q^2}{ 3 }  {\cal F} \left( \frac{1}{4}, \frac{1}{2}, \frac{5}{4}, -\frac{Q^2}{ b^2 r_{+}^4} \right) \right]
\end{equation}
When $ b \rightarrow \infty$, the above action approaches the Euclidean action for the RNAdS case which is computed in \cite{empa} as follows:
\begin{equation}
I_{RNAdS} = \frac{ \omega \beta } { 16 \pi G } \left( r_+ - \frac{ r_+^3}{l^2} - \frac{ Q^2}{r_+^2} \right)
\end{equation}

\section{Thermodynamics and Stability}
To discuss the thermodynamics and stability of the system one has to find a way to ``fix'' the potential since we are working with the grand canonical potential. However, due to the mathematical complexity of the action and the temperature, we will define a new variable $x$ as, 
\begin{equation}
x = \frac{Q}{ r_+^2}
\end{equation}
The reason for this substitution is the fact that the Hypergeometric function depends on $ Q^2/r^4$. Such a substitution does not change the results. Hence, the horizon $r_+$ as a function of $x$ and $\Phi$ becomes,
\begin{equation}
r_+ ( x, \Phi)  = \frac{ \Phi} { x  {\cal F}  ( \frac{1}{4}, \frac{1}{2}, \frac{ 5}{4}, -\frac{x^2}{b^2} ) }
\end{equation}
One can  write all the   thermodynamic quantities as functions of $x$ and $\Phi$. We will make  use of this substitution in computing all the thermodynamic quantities for fixed potential later in the section 4.

\subsection{Temperature}

When the black hole is extreme the temperature is zero leading to a diverging $\beta$. We varied the potential $\Phi$ and plot the $\beta$ against the horizon radii $r_+$. Following are the results:
\vspace{0.3cm}
\begin{center}
\scalebox{.9}{\includegraphics{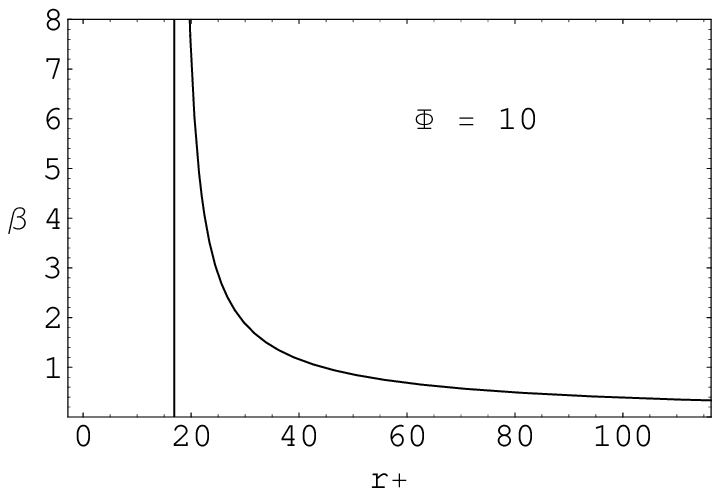}}
\vspace{0.3cm}
\end{center}
FIG.6: The inverse temperature $\beta$ vs. horizon radii $r_+$ at fixed potential $\Phi = 10$. Here $b = 2$, $ l=1  $, $ G =1$. 
\newline
\newline
In the Fig.6,  the temperature becomes zero leading to an extreme black hole. Hence at this potential the possibility of a RNAdS type black hole exists.

\vspace{0.3cm}
\begin{center}
\scalebox{.9}{\includegraphics{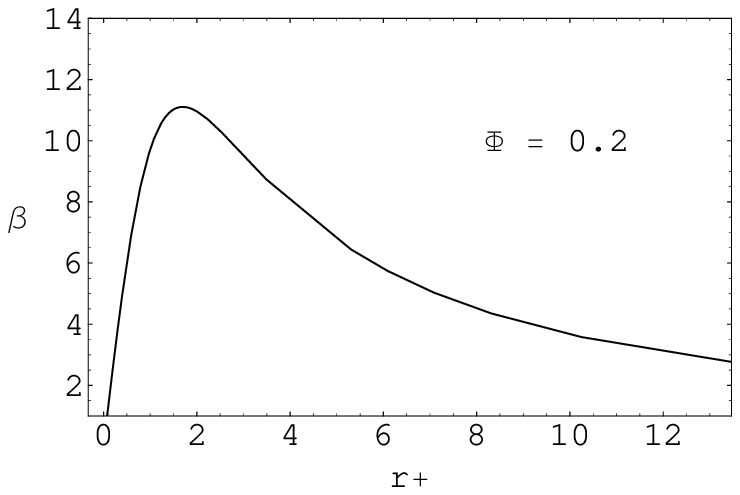}}
\vspace{0.3cm}
\end{center}
FIG.7: The inverse temperature $\beta$ vs. horizon radii $r_+$ at fixed potential $\Phi = 0.2$. Here $b = 2$, $ l=1  $, $ G =1$. 
\newline
\newline
In the Fig.7, the temperature does not become zero anywhere. Hence this may be  a SchAdS type or an RNAdS black hole. In fact for $\Phi=0$, $\beta$ behaves in a similar fashion as expected as given in Fig.8. \vspace{0.3cm}
\begin{center}
\scalebox{.9}{\includegraphics{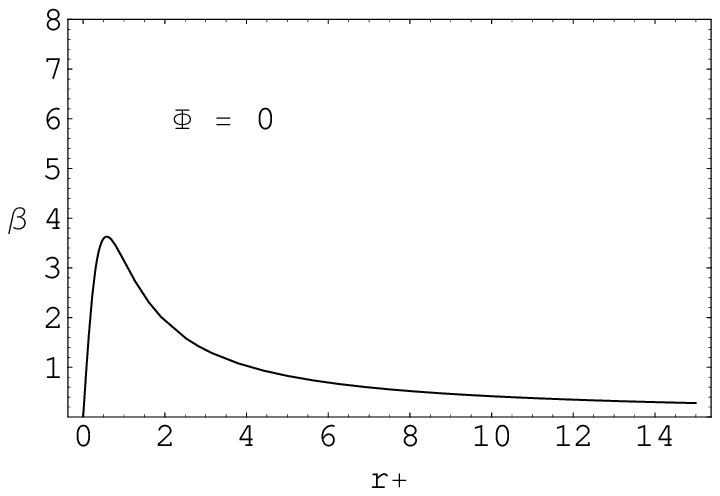}}
\vspace{0.3cm}
\end{center}
FIG.8: The inverse temperature $\beta$ vs. horizon radii $r_+$ at fixed potential $\Phi = 0$. Here  $ l=1  $ and $ G =1$. 
\newline
\newline
In comparison, the RNAdS black hole exhibits the exact same behavior when inverse temperature is considered against the horizon radii, which is discussed in \cite{empa}. The difference in the BIAdS case is that the Fig.7 could represent a black hole with two or one horizon.

\subsection{State Variables}
The state variables of the system may  be computed using the Euclidean action computed in section 3. Fist the entropy is computed as,
\begin{equation}
S = \beta \left ( \frac{ \partial I}{ \partial \beta} \right)_{\Phi} - I
\end{equation}
Note that this is computed as,
\begin{equation}
S = \beta \left ( \frac{ \partial I}{ \partial x} \right)_{\Phi} / \left ( \frac{ \partial \beta}{ \partial x} \right)_{\Phi} - I
\end{equation}
This expressions was shown to be equal to $ \frac{A} {4G}$ as expected. Here, $A$ is the area of the horizon of the black hole. The following figure represents the behavior of entropy with temperature for $\Phi = 10$.

\vspace{0.3cm}
\begin{center}
\scalebox{.9}{\includegraphics{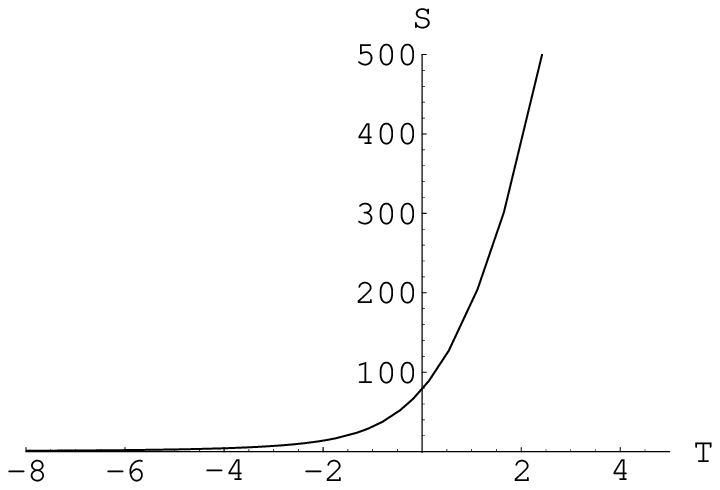}}
\vspace{0.3cm}
\end{center}
FIG.9: The entropy  $S$ vs. the temperature $T$ at fixed potential $\Phi = 10$. Here  $ l=1  $, $b =0.4$  and $ G =1$. 
\newline

The conserved charge $\tilde{Q}$ is given by,
\begin{equation}
\tilde{Q} = - \frac{1}{ \beta} \left( \frac{\partial I} {\partial \Phi} \right)_{\beta} 
\end{equation}
This expression was shown to be equal to $\frac{\omega Q}{ 4 \pi G}$ as expected \cite{empa}.

\subsection{Local Stability}
The local stability of the black holes can be studied by computing the specific heat  at constant potential:
\begin{equation}
C_{\Phi} = T \left( \frac{\partial S} { \partial T} \right)_{\Phi}
\end{equation}
By writing both $T$ and $S$ as functions of $x$ and $\Phi$ this can be computed symbolically easily. The exact expression is complicated and hence will not be written here. However, $C_{\Phi}$ will be plotted for various values of $\Phi$ as follows;

\vspace{0.3cm}
\begin{center}
\scalebox{.9}{\includegraphics{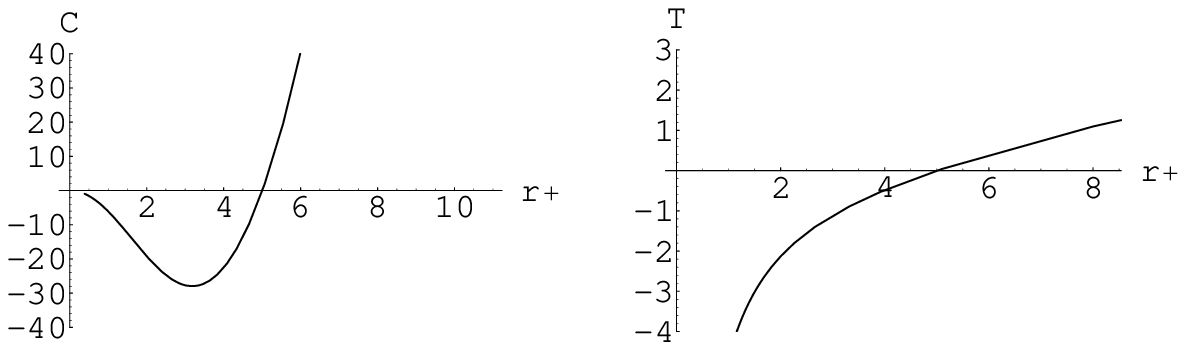}}
\vspace{0.3cm}
\end{center}
FIG.10: The specific heat  $C_{\Phi}$ and temperature $T$ vs. horizon radii $r_+$ at fixed potential $\Phi = 10$. Here  $ l=1  $, $b =0.4$  and $ G =1$. 
\newline

\vspace{0.3cm}
\begin{center}
\scalebox{.9}{\includegraphics{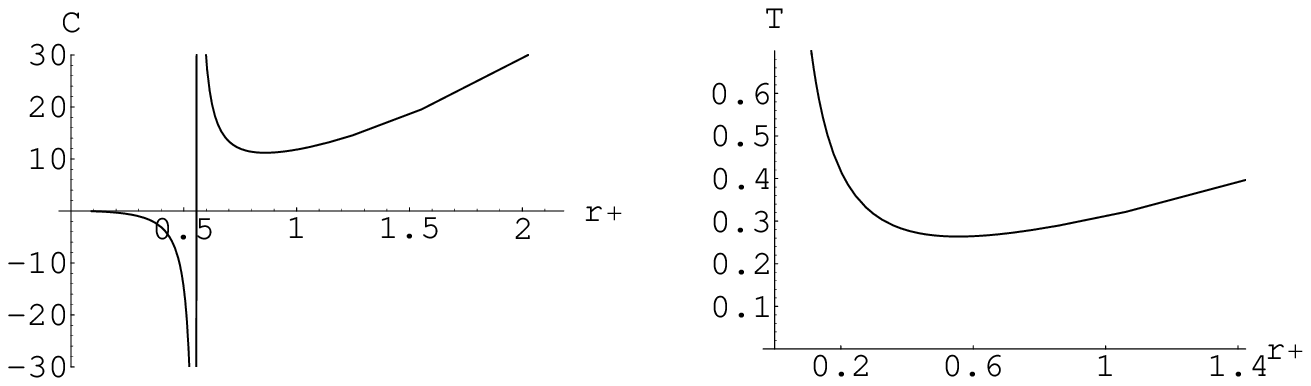}}
\vspace{0.3cm}
\end{center}

FIG.11: The specific heat  $C_{\Phi}$ and the temperature $T$ vs. horizon radii $r_+$ at fixed potential $\Phi = 0.3$. Here  $ l=1  $, $b =0.4$  and $ G =1$. 
\newline
\newline
The black holes represented in Fig.10 include extreme black holes since there is a horizon corresponding to zero temperature. The black holes for this potential is locally stable since the specific heat is positive for $ T > 0$. On the other hand, for non-extreme black holes given in the Fig.11, there are two branches to consider: small and large black holes. The large black holes have positive specific heat and are locally stable while the small black holes are unstable locally. There is a minimum temperature dividing these two branches.

\subsection{Gibbs  Free Energy and Hawking-Page Phase Transition}

  In order to study the global stability of the black holes, we will study the grand canonical (Gibbs) free energy of these solutions. The Gibbs free energy can be computer in terms of the Euclidean action as,

\begin{equation}
F =  \frac{I_E}{\beta}
\end{equation}
The free energy $F$ is plotted against the temperature $T$ for fixed potential in the following figure.

\vspace{0.3cm}
\begin{center}
\scalebox{.9}{\includegraphics{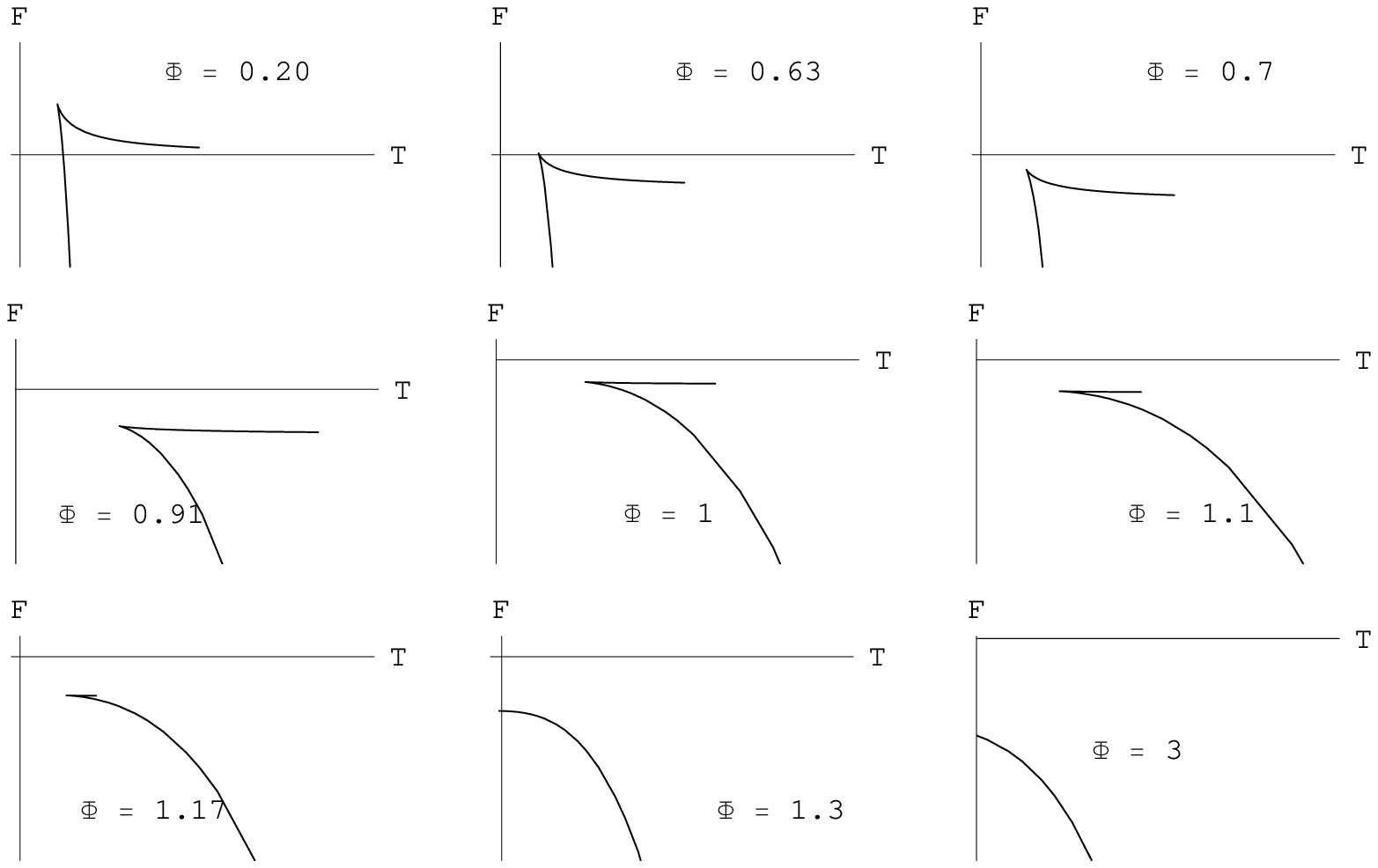}}
\vspace{0.3cm}
\end{center}
FIG.12: The free energy  $F$ vs. temperature $T$ at fixed potentials $\Phi = 0.2, 0.63, 0.7,0.91, 1, 
\newline
1.1, 1.17, 1.3, 3$. Here  $ l=1  $, $b =0.4$  and $ G =1$. The axis intersects at (0,0).
\newline
\newline
For small $\Phi$, there are no extreme black holes and hence the black holes don't reach zero temperature.  There are two branches for the small and large black holes. The large black hole branch has the smaller free energy. According to the Fig.12, for potentials smaller than $\Phi = 0.63$ (approximately), the pure-Ads geometry is preferred until the critical  temperature $T_{critical}$ is reached where the free energy of the larger black holes becomes zero. At this  critical  temperature there is a phase transition as discussed for SchAdS and is well known as Hawking-Page transition \cite{page}. Beyond the temperature $T_{critical}$, the large black holes have negative free energy and are preferred over pure-AdS geometry.

For $ \Phi >  0.63$, the phase transition occurs at the minimum temperature $T_{minimum}$ of the black holes. It is also interesting to note that there is a discontinuity in the free energy between the pure-AdS geometry and the black hole phase for potentials larger than 0.63. This behavior continues until  the potential reaches a critical value. Beyond that there is only one branch which include extreme black holes as represented by the $\Phi = 1.3$ case. Here for all positive temperatures, the black holes are globally preferred over the pure-Ads geometry. When one continues to increase the potential, the free energy continues to becomes smaller as represented by the $\Phi = 3$ case.

There is an interesting observation to make here in comparison with the RNAdS black hole phase transitions discussed in \cite{empa}. When there were two branches for the black hole free energy, the unstable smaller black holes always had positive free energy. In contrast, here, even the smaller branch does has negative free energy for some values of $\Phi$. There is a ``energy gap'' between the pure-AdS geometry and the black holes. The reason for this can be clarified as follows: In RNAdS black holes, at the critical potential where the extreme black holes start to appear, the maximum free energy is zero at $T =0$ point. On the other hand, for BIAdS black holes,  at the critical potential where the extreme black holes start to appear, the maximum free energy is lower than zero at $T=0$ point. This can be given by the following graphs,
\vspace{0.3cm}
\begin{center}
\scalebox{.9}{\includegraphics{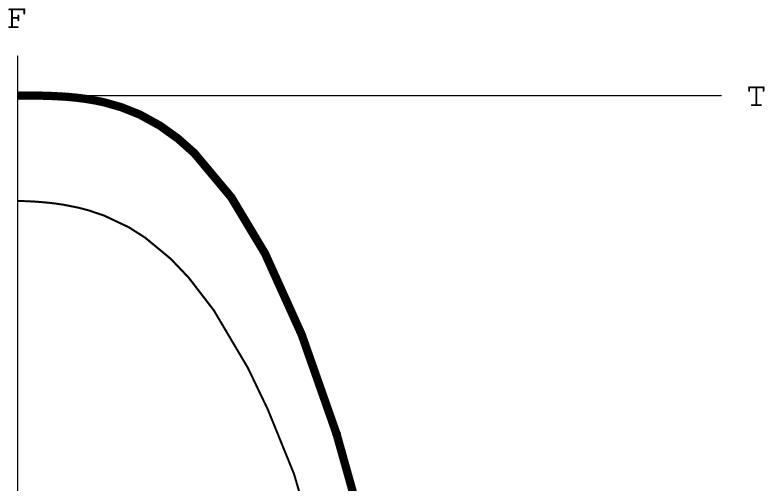}}
\vspace{0.3cm}
\end{center}
FIG.13: The free energy  $F$ vs. temperature $T$ for the RNAdS black hole and the BIAdS black hole at their respective critical potentials when the extreme black hole phase appear. The dark curve shows RNAdS black hole and the light curve shows the BIAdS black hole.
\newline
\newline
In the following figure, the free energy of the BIAdS black holes are compared to the RNAdS black holes for potentials leading to non-extreme black holes. The non-linear parameter $b$ is varied to observe how it effects the free energy.
From the graphs one can come to the conclusion that the BIAdS black hole is thermodynamically preferred over the RNAdS black hole.
\newpage
\vspace{0.3cm}
\begin{center}
\scalebox{.9}{\includegraphics{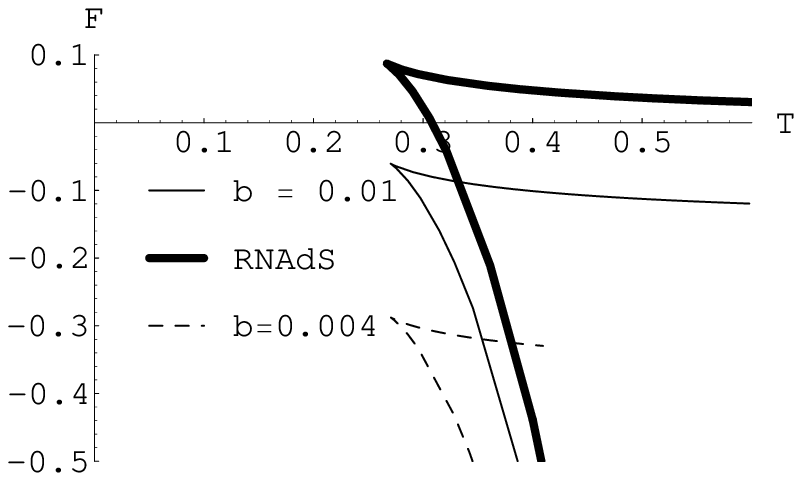}}
\vspace{0.3cm}
\end{center}
FIG.14: The free energy  $F$ vs. temperature $T$ at potentials $\Phi = 0.25$ for BIAdS and RNAdS black holes. Here  $ l=1  $ and $ G =1$.

\section{Conclusions}

In this paper we presented   some properties and thermodynamics of Born-Infeld black holes in the grand canonical ensemble.

BIAdS black holes represents properties similar to both SchAdS and RNAdS black holes. We present an interesting phase structure relating mass $M$ and the charge $Q$ of the solutions. The structure looks similar to the liquid-gas-solid phase diagram with $M$ representing $P$ and $Q$ representing temperature $T$. The ``Triple Point'' is at $M= \frac{1.31103}{b}$ and $Q= \frac{1}{2b}$. A natural question that comes to mind is to what kind of ``phase transition'' may occur when crossing the boundaries between SchAdS type black holes, RNAdS type black holes  and Naked Singularities. Also a question arises whether there is an analog ``Critical Point'' where the gas-liquid phase boundary ends in a  liquid-gas-solid phase diagram for the RNAdS black holes and Naked Singularities \cite{phase}. To answer such questions one has to 
study the  thermodynamics in the canonical ensemble. Since in a canonical ensemble the charge is fixed, it would be easier to relate thermodynamical behavior and phase structure between   $M$ and  $Q$ discussed in section 2. For the RNAdS black hole, the  free energy obtained in a canonical ensemble has similar behavior to a van der Waals-Maxwell liquid  gas system as described in \cite{empa}. It would be interesting to see if BIAdS black holes have similar properties.

In studying the thermodynamical stability of the BIAdS black holes, there are two cases to consider: extreme and non-extreme black holes. For the non-extreme black holes there are two branches consisting of small and large black holes. The smaller black holes are locally as well as globally unstable. On the other hand, the large black holes are locally stable since the specific heat is positive. Globally the large black holes are preferred for temperatures   $T > T_{critical}$ or $ T > T_{minimum}$ depending on the value of the electric potential $\Phi$. For temperatures smaller than $T_{critical}$ or $T_{minimum}$,  pure-AdS space is globally preferred.  There seems to be an ``energy gap'' between the pure-AdS geometry and the black holes for certain values of the potential. Further study should address the origin of this behavior. For the potentials consisting extreme black holes there is only one branch and the black holes are  locally as well as globally stable. 

There are several avenues to extend this work for one who is interested.

In an interesting paper, Gubser and Mitra \cite{gubser}, showed that there is a relation between thermodynamical instability and classical instability of large RNAdS black holes. Since the thermodynamical stability is well analyzed here for BIAdS black holes, one may study the classical stability under metric perturbations to see any similar behavior. BI black holes without the cosmological constant were shown to be stable under metric perturbations by the present author in \cite{fer3}. 

One of the possible extensions of this work is to study thermodynamics of the BIAdS black holes in $D=5$. The main motivation for this comes from the conjecture that physics of string theory in $AdS_5 \times S^5$ is  identical to $N=4$ gauge theory on the boundary of $AdS_5$. It has been tested for large SchAdS black holes 
in $D=5$  and has shown to match with the thermodynamics of the gauge theory in $D=4$ \cite{witten} \cite{suskind}.

The Born-Infeld black holes in de-Sitter space  would certain to offer new surprises. According to \cite{fer1} such BIdS black holes could behave similar to SchdS or RNdS black holes with maximum of three horizons. One wonder what phase structure it has between $M$  and $Q$. Since the possibility of three, two, one or no horizons, it should definitely be even richer than the one for BIAdS black holes. Furthermore, one can study the Hawking-Page phase transitions of BIdS black holes. Such phase transitions for RNdS black holes are discussed in \cite{carlip}.

Peca and Lemos \cite{lemos} studied the thermodynamics of the RNAdS black hole in the grand canonical ensemble using York's formalism \cite{york1} \cite{york2}. In this approach, the black hole is enclosed in a cavity with a finite radius to compete the action. The analysis of the BIAdS black hole in the York's formalism would be a worthy extension of the work presented here.


\begin{thebibliography}{99}


\bibitem{page} S. Hawking and D. Page, Comm. Math. Phys. {\bf 87} (1983) 577.

\bibitem{carlip} S. Carlip and S. Vaidya, Class. Quant. Grav. {\bf 20} (2203) 3827.

\bibitem{mal} J. Maldacena, Adv. Theor. Math. Phys. {\bf 2} 231 (1998) , hep-th/9711200.

\bibitem{witten} E. Witten, Adv. Theor. Math. Phys. {\bf 2} 505 (1998), hep-th/9803131.
\bibitem{born} M. Born and L. Infeld, Proc. Roy. Soc. Lond. {\bf A144} (1934) 425.
\bibitem{leigh} R. G. Leigh, Mod. Phys. Lett {\bf A4} (1989) 2767.
\bibitem{frad} E. S. Fradkin and A. A. Tseytlin, Phys. Lett. {\bf B163} (1985) 123.
\bibitem{tsey1} A. A. Tseytlin, Nucl. Phys. {\bf B276} (1986) 391.
\bibitem{gib1} G. W. Gibbons, hep-th/0106059
\bibitem{tsey2} A. A. Tseytlin, hep-th/9908105

\bibitem{sinclair} D.K. Sinclair and J.B. Kogut, hep-lat/0509097.

\bibitem{garcia} A. Garcia, H. Salazar and J.F. Plebanski, Nuovo.Cim {\bf 84} (1984) 65
\bibitem{demia} M. Demianski, Found. of Physics, {\bf 16} (1986) 187

\bibitem{nora1} N. Breton, gr-qc/0109022.

\bibitem{nora2} N. Breton, Phys.Rev. {\bf D67}  (2003) 124004

\bibitem{gib2} G. W. Gibbons and C. A. R. Herdeiro, Class. Quant. Grav. {\bf 18} (2001) 1677

\bibitem{tamaki1} T. Tamaki and T. Toriii, Phys.Rev. {\bf D62}  (2000) 061501.


\bibitem{clem} G. Clement and D. Gal'tsov,  Phys.Rev. {\bf D62} (2000) 124013.

\bibitem{tamaki2} T. Tamaki, JCAP {\bf 0405} (2004) 004



\bibitem{fer1} S. Fernando  and  D. Krug, Gen. Rel. Grav. {\bf 35} (2003) 129

\bibitem{cai} R.  Cai, D. Pang and A. Wang,  Phys. Rev. {\bf D70} (2004) 124034

\bibitem{dey} T. K. Dey, Phys.Lett. {\bf B595}  (2004) 484.



\bibitem{gibbons4} G.W. Gibbons and S.W. Hawking, Phys. Rev. {\bf D 15} (1977) 2752.


\bibitem{ras} D. A. Rasheed, hep-th/9702087.



\bibitem{abra} M. Abramowitz and I.A. Stegun, ``Handbook of Mathematical Functions '', Dover, 1972.
\bibitem{nora3} N. Breton, Phys. Rev.  {\bf D72} (2005)  044015.

\bibitem{phase} N. Goldenfeld, ``Lectures on Phase transitions and Renormalization Group '', Addison Wesley, 1992.


\bibitem{empa} A. Chamblin, R. Emparan, C.V. Johnson and R.C. Myers, Phys.Rev. {\bf D60} (1999) 064018.



\bibitem{bala} V. Balasubramanium and P. Kraus, Comm.Math.Phys. {\bf 208} (1999) 413.

\bibitem{mann} R.B. Mann, Phys. Rev. {\bf D 60} (1999) 104047.

\bibitem{empa2} R. Emparan, C.V. Johnson and R. C. Myers, Phys.Rev. {\bf D60} (1999) 104001.

\bibitem{rod}  R. Olea, JHEP {\bf 0506} (2005) 023.

\bibitem{gibbons2} G.W. Gibbons, M.J. Perry and C.N. Pope, Class.Quant.Grav. {\bf 22} (2005) 1503





\bibitem{gubser} S. S. Gubser and I. Mitra, hep-th/0009126

\bibitem{fer3}  S. Fernando, Gen.Rel.Grav. {\bf 37}, (2005) 585.

\bibitem{suskind} L. Susskind and E. Witten, hep-th/9805114


\bibitem{lemos} C.S. Peca and J.P.S. Lemos, Phys. Rev.  {\bf D 59} (1999) 124007. 

\bibitem{york1} J.W. York, Phys. Rev. {\bf D 33} (1986) 2091.

\bibitem{york2} H.W. Braden, J.D. Brown, B.F. Whiting and J.W. York, Phys. Rev. {\bf D 42} (1990) 3376.






\end{thebibliography}
\end{document}